# Evidence for Topological Band-to-Band Transitions in a Type-II Weyl Semimetal

Seyyedesadaf Pournia [a], Giriraj Jnawali [a], Samuel Linser [a], Howard E. Jackson [a], Leigh M. Smith [a]

[a] Department of Physics, University of Cincinnati, Cincinnati, OH, USA

## Abstract:

The ternary van der Waals material $NbIrTe_4$ is a Type-II Weyl semimetal. We use a tunable circularly polarized mid-infrared laser to investigate the existence of band-to-band excitations using transient reflectivity in an exfoliated nanoflake and photothermoelectric and photogalvanic signals in a device. Unpolarized photothermoelectric spectroscopy shows that the absorption in the Weyl semimetal increases rapidly above 0.3 eV as expected from the increase in the density of states from DFT calculations. However, the reflectivity shows a sign change in the circular dichroism of the reflected light which is dominated by $\sigma$- light for energies below 0.5 eV and $\sigma$+ light for energies above that energy. Using an intense pulse to perturb the electrons near the Fermi level, we show that this 0.5 eV energy is associated with a band-to-band transition from a band slightly below the Fermi energy to a higher lying empty band. We use spectroscopy of the circular photogalvanic effect (CPGE) from 0.3 to 1 eV to show a strong peak near 0.5 eV which lies on top of a two orders-of-magnitude increase at the lowest energies as the Weyl points are approached. We conclude that this band-to-band optical transition enhances the Berry Curvature responsible for the CPGE, and may involve the Weyl points just below the Fermi energy.

**Introduction:**

Semimetals are characterized by a very weak density of states near the Fermi level which is caused by touching of the conduction and valence bands at isolated parts of the Brillouin zone. For many materials such band inversions and crossings result in a wide variety of topological features in the electronic structure which are associated with a topological invariant (eg. Chern number) and low energy Fermionic excitations, such as Dirac or Weyl fermions, which are analogues to particles in high energy quantum field theory.[1] Such materials have been under intense theoretical and experimental investigations over the past decade yielding the discovery of a large number of unexpected topological structures which have been characterized by the dimensionality (0, 1 or 2 dimensions), degeneracy (2, 3, 4, 6 or 8 fold degeneracies), topological charges (invariant), band dispersions of the crossings (Type I or II), and the crystal symmetries which stabilize them.[1] Distinct signatures of these topological excitations have been seen in surface states (Fermi arcs), unusual transport features, and linear and nonlinear optical effects. Now that such topological electronic structure has been confirmed, a critical challenge has been to find how these features interact with trivial (non-topological) excitations associated with the electron or phonon band structures.

   Weyl semimetals (WSM) are a group of three dimensional extraordinary topological materials which exhibits linear energy dispersion in the vicinity of the points where the bands are crossing. The carriers responsible for the low energy excitations close to the crossing points are massless fermions.[2,3,4,5,6] The fermions in a Weyl semimetal have a definite chirality, either -1 or +1, which are separated in momentum space, making them stable against small perturbations.[7] As the nodes are assumed to be like magnetic monopoles in momentum space, they always exist in pairs. The projection of the nodes with opposite chirality on the Fermi surface are connected via open surface states known as Fermi arcs.[8] Direction detection of these surface arcs is a signature of the existence of the Weyl nodes and were first detected in the type I Weyl semimetal TaAs, and in a number of other systems.[9,10,11,12,13,14] Despite this, the chirality of carriers were confirmed via exploring circular polarization-dependent currents in transport experiments where Fermions were excited between the branches making up the massless linear crossings of the Weyl points.[15,16,17] Many such materials have shown strong effects in temperature, excitation or field-dependent phonon Raman scattering measurements which suggest that the topological electronic structures have strong effects on the phonons through electron-phonon coupling. There have been almost no demonstrations thus far on the impact of these features on trivial (non-topological) electronic states.

   Optical transitions in materials promote electrons from states which lie below the Fermi energy and so are occupied, to unoccupied states which lie above the Fermi energy. Absorption and reflectivity have been used to identify states close to the band edges in gapped systems such as the topological insulator Bi2Se3 or the chiral semiconductor Tellurium which has unoccupied Weyl nodes in the conduction band. However, it is extremely difficult in gapless semimetal systems to unambiguously identify the initial and final states in an optical transition of photon energy, E, because there can be many possible transitions ranging from initial states below the Fermi level by the photon energy, E, to final states which are above the Fermi level by the photon energy, E. Linear polarization effects can help to limit the number of possible states, but it is important to realize that linear dichroism is naturally allowed in non-magnetic Weyl semimetals because inversion symmetry is broken. On the other hand, circular dichroism is only seen when time-reversal symmetry is broken (not the case for non-magnetic Weyl semimetals), or it is allowed by crystal symmetry. The Weyl nodes in particular are expected to show strong circular polarization effects.

In this paper, we use the response of the Type-II Weyl semimetal $NbIrTe_4$ to circularly polarized mid-infrared light to identify potential band-to-band transitions associated with the chiral Weyl points. $NbIrTe_4$ was predicted to be a type II Weyl semimetal from first-principal band structure calculations, hosting 8 pairs of Weyl nodes in the first Brillouin zone.[18] This has been confirmed through ARPES visualization of the surface Fermi arcs,[19] with recent observations of non-saturating magneto-resistance,[20,21] quasiparticle interference observed in magneto transport and STM spectroscopy,[20–22] and strong electron-phonon effects seen in resonant Raman scattering and STM spectroscopy.[22,23] It crystalizes in a non-centrosymmetric orthorhombic Bravais lattice belonging to the $Pmn2_1$ space group. *Ab initio* DFT calculations of the band structure have shown that the linear crossings for all Weyl points extend only a maximum of 100 meV (or less), and so our measurements which are made from 0.3 to 1.0 eV are all well above any energies which involve interband transitions between the Weyl points linear bands.

**Experiment and discussion:**

A single crystal of NbIrTe4 was grown by the flux method (see Methods).[23] Nanoflakes were exfoliated and deposited onto a silicon substrate which has a 300 nm SiO2 insulating layer. Most of the nanoflakes were 80 to 100 nm thick and 5 to 10 μm wide and 10 to 20 μm long. We also fabricated a two terminal device from a similar $NbIrTe_4$ nanoflake (figure 1.c). In the device the crystal was aligned between the contacts so the current flowing in the device is along the a axis. As $NbIrTe_4$ is an orthorhombic crystal, the b-axis is perpendicular to the current flow. The current in the device was controlled by an SRS CS580 current source and the photovoltage induced by laser excitation was amplified and then measured with a lock-in technique. The lasers used in the photovoltage measurement was a 200 fs pulsed Ti:Sapphire laser which could be tuned from 680 -1030 nm (1.8 to 1 eV). This 820 nm output also pumped an OPO laser which could be tuned between 1030 - 4000 nm (1 to 0.3 eV). The laser was focused onto the device using a 40 x 0.5 N.A. reflective objective at normal incidence with respect to the surface of the sample.

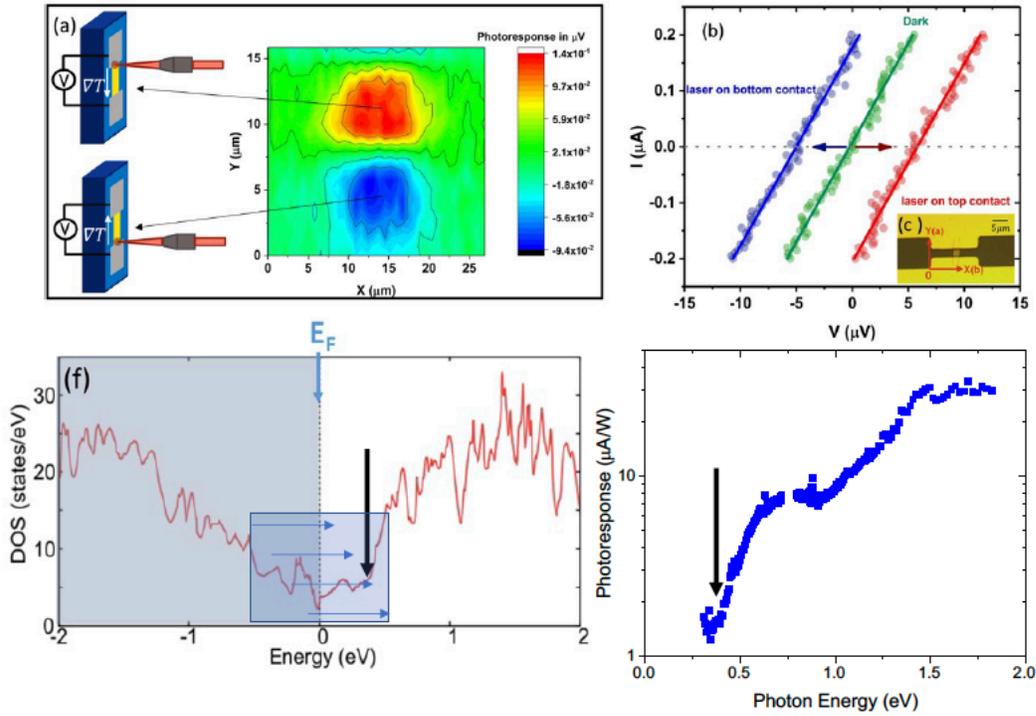

Figure 1. (a) Photovoltage spatial map of device excited at 1540 nm. Schematic diagrams showing phototermoelectric response at top and bottom contact. (b) I-V measurements: in the dark (green), 1540 nm laser focused close to the bottom contact (blue) and top contact (red) of the device shown in (c). (c) Photoresponse spectra of the device which is dominated by the photothermoelectric effect. (d) NbIrTe4 Density of States (DOS) from DFT calculation. The grey box notes range of possible initial and final states for a 0.5 eV optical transition (see horizontal arrows).

## Photothermoelectric Spectrum of Absorption

Figure 1.a shows the photoresponse of the device to 1540 nm laser light which was linearly polarized along the a-axis of the crystal. The device was placed in a chip carrier which was mounted to the cold finger of the cryostat. Using piezo-manipulators, the sample was raster-scanned along the x and y directions, with the laser spot fixed. The origin and axes of this mapping is marked in figure 1.c. In this measurement the device current was zero (open circuit conditions). The mapping shows a maximum photovoltage when the laser is focused near either of metal contacts, while the central midpoint between the metal pads exhibited zero signal. This behavior is consistent with the photothermoelectric effect (PTE) which has been seen in many materials including topological insulators, Dirac and Weyl semimetals[24] [25]. When the laser is focused close to one contact, the power absorbed induces a temperature gradient between the two ends of the device and the hot carriers will flow toward the other contact. When the laser moves to the midpoint between the contacts the induced photovoltage vanishes. As the laser focuses on the other contact, the temperature gradient and electron flow reverse direction. As a consequence, the photothermoelectric effect induces opposite electric fields or electric currents close to the two metal contacts. In figure 1.b the current versus voltage measurements in the device are measured in the dark (green), under 1540 nm laser illumination while the laser is focused near the top contact (red), and when the laser was focused near the bottom contact (blue). All measurements were made at 300K. All three profiles are linear, indicating the contacts are ohmic. The slopes do not change, indicating that the conductivity of the device is not changed under illumination. The x-intercepts are on opposite sides of zero showing that focusing the laser on each contact causes a shift to opposite biases. To make this shift visible, the two data sets under the laser illumination are rigidly shifted by 35 times the induced voltage at 0 current for each of them. The resistance in the dark or under illumination is seen to be a constant 27 Ohms, which suggests a resistivity in the device to be 1.87 $\Omega.\mu$m, which is in good agreement with reference [20]. By considering the mobility of the electrons to be 5 cm$^2$/V.s (see Ref. [26]), we estimate the room temperature density of carriers to be 6.68 x 10$^{21}$ cm$^{-3}$.

Using the photothermoelectric effect it is possible to measure the absorption as a function of energy.[20,21] To obtain such a photothermoelectric spectrum the laser is focused on the point fixed close to the metal contact with the maximum positive photoresponse and the energy is tuned over the entire range of energies. The photovoltage is normalized by the power of the laser at every excitation energy. The photovoltage was divided by the resistance of the device to interpret the photosensitivity in terms of the more familiar current ($\mu$A/W) instead of voltage. All measurements were made under open circuit conditions (I=0). Figure 1.e shows the signal for the unpolarized laser as a function of photon energy from 0.31 eV (4000nm) to 1.82 eV (680nm). The signal display a strong onset in the absorption at about 0.4 eV, originating from a significant increase in the DOS at that energy, which is confirmed by the DFT calculation in figure 1.f.

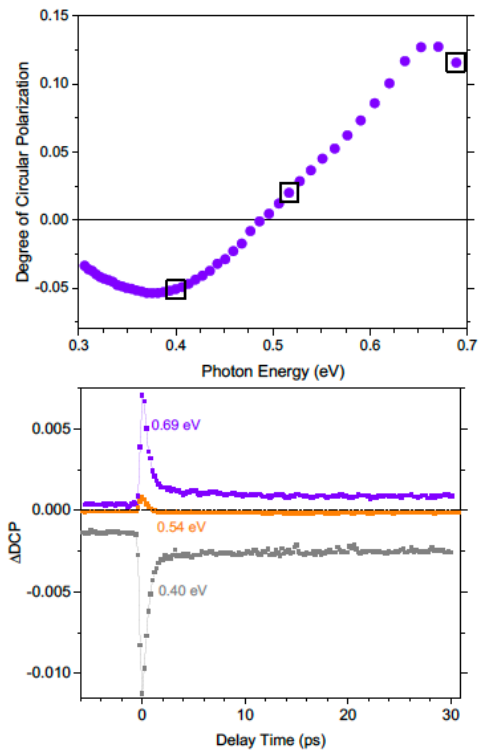

Fig. 2. (a) Degree of circular polarization (DCP = (I+ - I-)/(I+ + I-)) as a function of excitation energy. (b). Transient reflectivity of the change in DCP (ΔDCP) resulting from excitation by a 1.5 eV pump pulse at 0.69, 0.54 and 0.4 eV probe pulses.

**Circular Dichroism for Reflected Light**

As we noted previously, it is extremely difficult to associate a particular absorption onset with a particular set of bands. The bands involved could range from 0.4 eV below the Fermi energy to a band 0.4 eV above the Fermi energy. To try and narrow the range of possibilities, we measure the circular dichroism of reflected light. It is important to note that for normal incidence at zero magnetic field, NbIrTe4 which has $Pmn2_1$ crystal symmetry should not be sensitive to circular polarized light.[27] On the other hand, the Weyl points should in fact be sensitive to circular polarized light. We use a ZnSe photoelastic modulator to vary the polarization from σ+ to σ- at a 50 kHz frequency of the incident light as it is tuned from 0.3 to 0.8 eV. The reflected signal is detected by an $LN_2$-cooled InSb detector and measured by a lockin amplifier and normalized by the incident laser power. In Fig. 2 (a), the degree of circular polarization (DCP) signal (DCP = $(I_+ - I_-)/(I_+ + I_-)$) ranges from +10 to -10% and there is a clear sign change at 0.5 eV so that the reflected light is dominated by σ- polarized light at lower energies and σ+ polarized light at higher energies. While this does not reduce the number of bands which potentially could be involved, if it is possible to calculate the optical response for band-to-band transitions it potentially could be used to narrow substantially the range of possibilities.

**Transient Circularly Polarized Reflectivity**

To narrow substantially the range of bands which can be involved, we use transient polarized reflectivity. A 1.5 eV intense pump pulse is used to excite the nanoflake, and then the change in the DCP signal is measured using a second lockin which is referenced to an optical chopper which modulates the pump beam. The electrons which are excited to very high energies rapidly scatter down to the Fermi energy within several picoseconds. This causes a distribution of very hot (~600 K) electrons at the Fermi energy. Because the *only* change in the electron distribution occurs near the Fermi energy, the *only* modulation of the circular dichroism signal occurs for optical transitions where the initial or final states must be within $\pm k_B T$ of the Fermi energy. Fig. 2 (b) shows time decays taken at 0.7 eV, 0.54 eV and at 0.4 eV. Not surprisingly, the dynamic signal at 0.54 eV (near the crossing of the dichroic reflected signal shown in Fig. X) is extremely weak. More importantly, a clear sign change in the dynamic circular dichroism reflectance is seen: positive at energies above 0.55 eV and negative at lower energies. This suggests strongly that there is a band-to-band transition near 0.55 eV, and also that this transition *must* involve states which are close to the Fermi energy. The notable degree of circular polarization (DCP) which is observed both for reflectivity of mid-IR light and also in pump-probe transient reflectivity is striking. We develop here a simple heuristic model which strongly suggest a direct connection of the DCP experimental results with specific electronic states.

DFT calculations show four Weyl points located in each of the $k_z = \pm 0.2$ planes, with linear crossings located approximately 80 meV below the Fermi energy (Fig. 3(a)). These linear crossings are not Lorentz invariant but exhibit different velocities: $E = -v_l k$ and $E = +v_u k$ where $E = 0$ is located at the crossing point. We assume that the chirality is $\chi = +1$ and the branch with $-v_l$ velocity is spin down and the branch with $v_u$ velocity is spin up (Fig. 3(a)). We consider optical transitions which promote electrons from the Weyl points to an empty Kramers degenerate parabolic band at $+E_0$ above the Weyl crossing (well above the Fermi energy). Right circularly polarized light, $\sigma+$, of energy $E+$ can thus only promote a spin down electron from the $-v_l$ branch to the spin up states in the parabolic band (see Fig. 3(a)). Left circularly polarized light, $\sigma-$, of energy $E-$ can only promote a spin up electron from the $+v_u$ branch to the spin down states in the parabolic band. We assume that momentum is conserved in the transition so the electrons only change energy. The energies for these transitions as a function of momentum can be written as $E+ = E_0 - v_l k + k^2/2m$ and $E- = E_0 + v_u k + k^2/2m$.

Optical transitions which conserve momentum must promote electrons from bands which are occupied to states at higher energy which are unoccupied. From Lasher and Stern[28] it is possible to relate the absorption for such a process to the stimulated emission:

$$r_{stim+} = \frac{4\,n\,e^2 E_+}{m^2 \hbar^2 c^3} |M_+|^2 \rho_{red+}(E_+)(f_u - f_l) \qquad (1)$$

$$r_{stim-} = \frac{4\,n\,e^2 E_-}{m^2 \hbar^2 c^3} |M_-|^2 \rho_{red-}(E_-)(f_u - f_l) \qquad (2)$$

We assume that the matrix elements, $M_+$ or $M_-$, are constant for any allowed excitation for $\sigma+$ or $\sigma-$ light. We assume that the parabolic band is empty (well above the Fermi level) while the linear bands are partially occupied. Fermi-Dirac functions thus can relate these occupations as $f_u = 0$ and $f_l = f_\pm = \frac{1}{1 + e^{(E_\pm - E_0 - E_{Fermi})/k_B T}}$ (see Fig. 3(b)). The reduced density of states can be calculated as:

$$\rho_{red\pm} = \frac{k^2}{2\pi^2} \frac{dE_\pm}{dk} \qquad (3)$$

and so the absorption for $\sigma+$ or $\sigma-$ light with energy E becomes

$$\alpha_\pm(E) = -\frac{\pi^2 c^2 \hbar^3}{n^2 E^2} r_{stim\pm}(E) \qquad (4)$$

Because Weyl crossings are just below the Fermi energy, the absorption depends sensitively on the electronic temperature. The inset to Fig. 3(b) shows the density of states for the linear bands and the parabolic band along with the associated Fermi-Dirac distributions at 300 K and 600 K. The inset to Fig. 3(c) shows the energy dependent absorptions for left ($\alpha-$) and right ($\alpha+$) circularly polarized light at 300 K and 600 K. By subtracting these two absorptions it is possible to calculate $\alpha_{CP} = \alpha_+ - \alpha_-$ at each electronic temperatrue. One can immediately draw several conclusions. First, because the DOS is zero at the crossing point, the absorption is zero at $E_0$ (at k = 0.). Secondly, it is only possible to obtain a *non-zero* $\alpha_{CP}$ if the linear bands are non-lorentzian (tilted). A transient reflectivity measurements can *only* measure a change in the reflectivity if there is a modulation $\Delta\alpha_{CP}$ in the absorption caused by the hot electrons at the Fermi level caused by the pump pulse. Fig. 3(c) shows that $\Delta\alpha_{CP} = \alpha_{CP}(T^H_e) - \alpha_{CP}(300 \text{ K})$ is negative below and positive above 0.5 eV. Note that these absorptions depend on the electronic temperature with the amplitude of the signal becoming smaller as the electronic temperature relaxes from 600 K down to the lattice temperature of 300 K. These results quite closely reflect the transient reflectivity decays shown in Fig. 2(c).

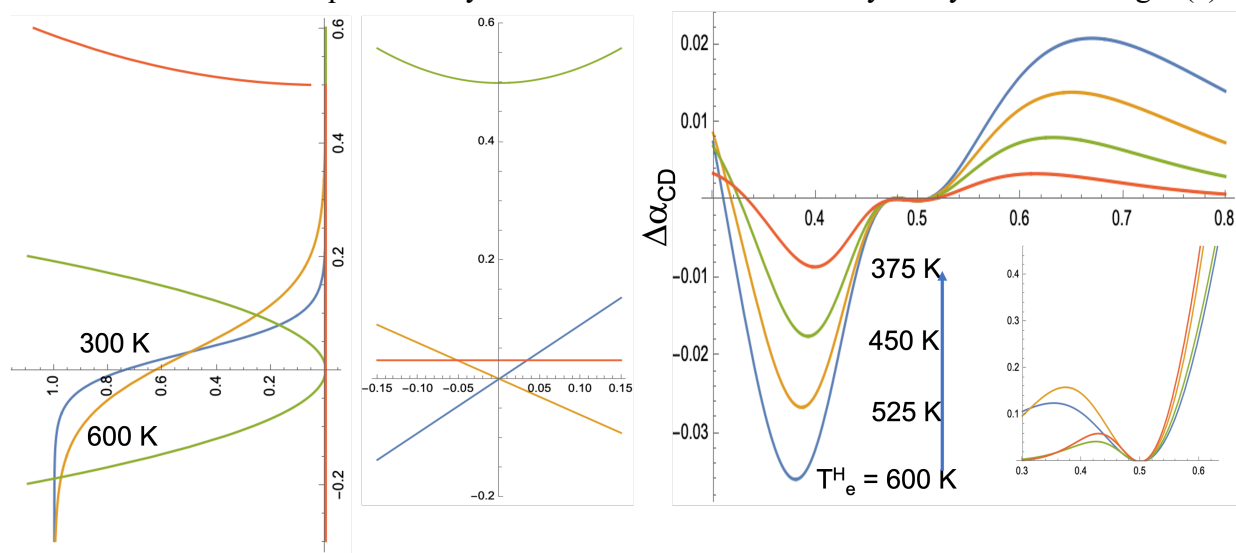

Fig. 3. (a) Simple model of interband optical transitions between Weyl crossings and Kramers degenerate parabolic higher lying band. Fermi level is 0.03 eV above Weyl crossing point. Linear bands are not Lorentz invariant but have different velocities. Optical transitions for s+ and s- polarized light shown by arrows. (b) DOS for Weyl point and Kramers degenerate band. Fermi-Dirac distribution for electron temperatures of 300 K and 600 K. (c) Inset shows optical absorption for s+ and s- polarized light at 300 K and 600 K. Change in CP absorption due to electronic temperature change from 300 K to 600 K: ($\Delta\alpha_{CD}$ = ($\alpha$+(600K) – $\alpha$-(600 K)) – ($\alpha$+(300 K) – $\alpha$-(300 K)). The calculation shows an increased $\alpha$- absorption at energies below 0.5 eV and increased $\alpha$+ absorption at higher energies.

It is important to note that only the electronic temperature at the Fermi level changes and not the lattice temperature. This dynamic hot electron relaxation process is monitored by a delayed mid-IR probe pulse. The reflectivity is measured as the difference in the reflected signal with the pump on versus the pump off. Because the reflectivity *only* changes because of the hot carriers which are at the Fermi level, this measurement only probed initial or final states which are close to the Fermi level.

While this simple model is calculated for optical transitions between a Weyl crossing and a higher lying parabolic band, the result is very general. Because a transient reflectivity measurement probes *only* optical transitions which are impacted by the electronic temperature (see Fig. 3(b)), *only* bands close (~ ± 100 meV) to the Fermi energy contribute to the *change* in the reflectivity. As noted by others, there are only *four* bands which cross the Fermi level in NbIrTe$_4$ and so transient reflectivity measurements are able to probe optical transitions associated with these bands with extremely high sensitivity.[20,21]

**CPGE Measurement:**

All of the measurements described thus far are a linear response of the material to light and so are a direct measure of the dielectric response. It has been shown that in crystals which lack a spatial center of symmetry, a photocurrent is induced when the sample is illuminated by polarized light even in the absence of external electric field.[29–31] This nonlinear response to the electric field of the light is called the photogalvanic effect (PGE) and can be divided into two parts: the photoresponse to 1) linear and 2) circular polarization.[30,31] In the presence of time-reversal symmetry the Linear Photogalvanic Effect (LPGE) is equivalent to the shift current which is caused by the electron-hole polarization in the presence of the oscillating electric field and so can be related to the difference between the Berry connections (the average real space position of the carrier) for the electron and hole.[30,31] The Circular Photogalvanic Effect (CPGE) is equivalent to the injection current which can be directly related to the Berry Curvature. This nonlinear effect combines two oscillating electric fields and rectifies them creating a DC current which depends on the circular polarization. Because the Berry curvature probes the geometrical properties of the wavefunction it is sensitive to the topology of the states. If the CPGE is enhanced approximately at the same energies where the circular dichroism for reflected light goes to zero, and where the transient response goes to zero, then this is strong indication that the interband transition is associated with topologically non-nontrivial electronic states.

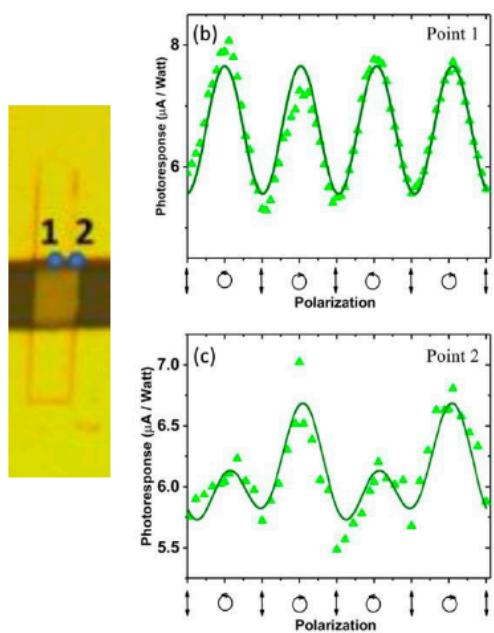

Fig. 4. (a) Optical image of device with point 1 at the center of the nanoflake close to the top contact, and point 2 at the edge of the nanoflake close to the top contact. (b). Photoresponse of the device excited at point 1 as a function of λ/4 plate orientation with left and right circularly polarized excitation marked. No CPGE is observed. (c). Photoresponse of the device excited at point 2 as a function of λ/4 plate orientation showing strong CPGE response.

To measure the CPGE response, we use a full-wave liquid crystal (LC) variable wave plate. The LC drive voltage is set to a quarter wavelength for each specific probe wavelength, and by rotating the crystal for 360°, we can measure the photoresponse from linear polarization (along the current direction and the <span style="color:red">a</span> crystal axis), circular polarization (right and left handed), and the elliptical polarizations produced in between. As shown in Fig. 4, for the laser focused on the contact near the center of the flake, there is no change in the current for left or right circularly polarized light. However, if the laser is focused on the edge of the nanoflake then one observes a significant difference in the current for left or right circularly polarized light.

We note that the second order injection current for CPGE can be written as:

$$j_i(t) = i\beta_{ij}(\boldsymbol{E} \times \boldsymbol{E}^*)_j \qquad (5)$$

This current is zero for linearly polarized light, but is nonzero and changes sign for left and right circularly polarized light if $\beta_{ij}$ is nonzero, where $\beta_{ij}$ is related to the Berry curvature. Because of the mirror and rotational symmetries of the NbIrTe4 lattice the CPGE tensor $\beta_{ij}$ cannot generate an injection current at normal incidence which explains the absence of CPGE when the laser is centered on the nanoflake. However, at the edge of the nanoflake, the crystal symmetries are broken perhaps explaining the presence of the CPGE effect.[32] It is also known that for the laser *focused* onto the nanoflake, that the *spatially dispersive* CPGE signal is present which causes rotational currents around the laser spot center.[15] At the center of the flake, these currents going into and out of the contact should cancel. At the edge of the flake, only the rotational current inside the edge contributes, which also might also explain the measured CPGE signal.

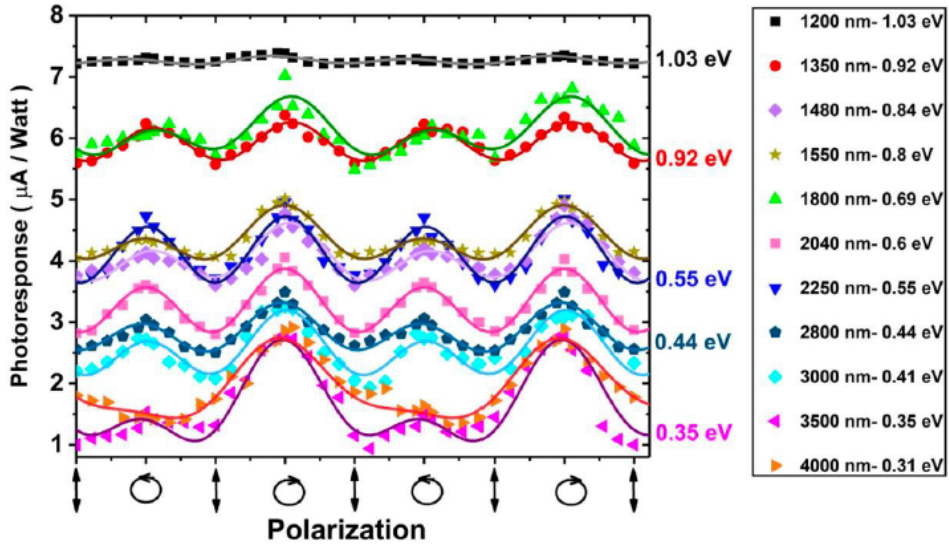

Fig. 5. Energy dependence of the photoresponse excited at point 2 (edge of device) as a function of λ/4 plate orientation. The PTE response is seen to decrease rapidly as the excitation energy decreases, while the CPGE increase substantially over the same energy range. Lines are fits to $I = C \sin(2\theta) + L \cos 4(\theta + \varphi) + D$.

In either case, we can now measure how the CPGE strength changes as a function of energy. Fig. 5 shows the induced current as a function of quarter waveplate angle for 11 different wavelengths from 1200 nm to 4000 nm (0.3 to 1.0 eV). These results at each energy were fit to the following equation for CPGE:

$$I = C \sin(2\theta) + L \cos 4(\theta + \varphi) + D \qquad (6)$$

The first term on the right side is due to the circular photogalvanic effect (CPGE) which shows that right- and left-handed circular produces equal amplitude electric currents (C) but in opposite directions. The second term (L) comes from the linear photogalvanic effect which originates from the symmetric part of the second order response to light. The third term (D) is the photothermoelectric effect (PTE). The result of this fitting for the parameters C and D as a function of excitation energy is shown in Fig. 6. This result clearly shows that the PTE decreases by an order of magnitude with decreasing photon energy. In contrast, the CPGE term increases by nearly two orders of magnitude at lower photon energies, but also shows a significant peak from 0.5-0.8 eV. (The LPGE (L) term is nearly constant with energy and is not shown here.)

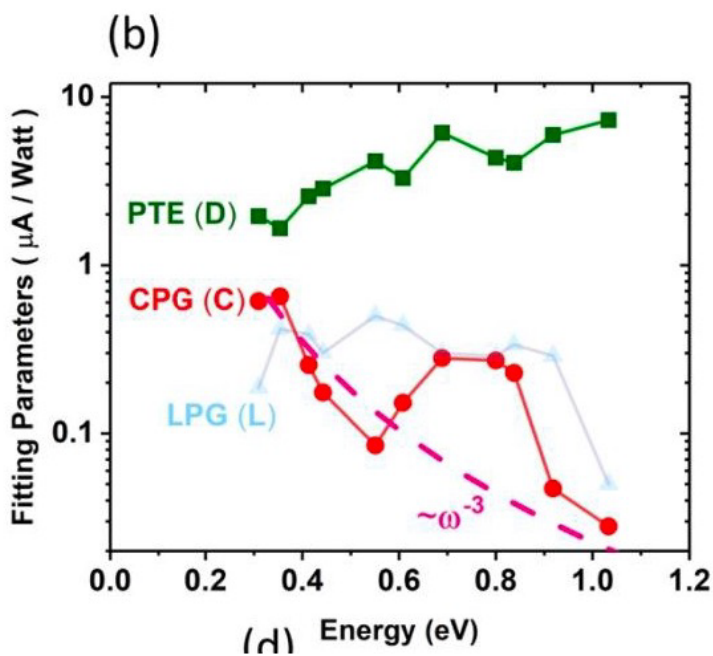

Fig. 6. Energy dependence of the CPGE response (C) and the PTE response (D) as a function of energy (0.3 to 1.0 eV) on a log scale. The PTE response is seen to decrease by an order of magnitude at lower energies while the CPGE response is seen to increase by two orders of magnitude. A strong peak is seen from 0.5 to 0.8 eV.

These three experimental results provide strong evidence that there is a direct band-to-band optical transition near 0.5 eV which contributes to the CPGE. First, the circular dichroism (CD) of the reflected light goes through a zero near 0.5 eV. Secondly, circularly polarized transient reflectivity measurements show a sign change in the dynamic CD also near 0.5 eV. Because the 1.5 eV pump pulse can only modify the hot electrons near the Fermi energy, this means that either the initial or final state in the optical transition *must* be within 100 meV of the Fermi energy. We develop a simple model to show that the circular polarized transient reflectivity is explained by an optical transition between a Weyl crossing just below the Fermi energy and a higher lying Kramers degenerate band. Finally, the strength of the CPGE in a device is seen to substantially increase in a range (0.5 to 0.8 eV) just above this 0.5 eV transition. This suggests that the presence of the band-to-band optical transition contributes to the Berry Curvature which increases the CPGE in that energy range.

**Conclusion:**

Taken together the results here describe a new way of extracting band-to-band transitions which are in topological semimetals. As noted previously, because semimetals do not have a gap, it is extremely difficult to disentangle the numerous optical transitions which can occur in general for a particular photon energy. However, because transient reflectivity (or absorption) pump probe measurements only modify the electronic distribution at the Fermi level, evidence of transitions in polarized dynamic measurements can provide a detailed insight into the nature of those transitions which involve electronic bands which are close to the Fermi level. The number of such electronic states in semimetals are by their nature small. By combining with nonlinear measurements, such as CPGE, one can also determine whether these band-to-band transitions are between trivial bands or exhibit exhibit a geometrical quantity such as the Berry curvature. Understanding the nature of such transitions will require more detailed band structure calculations and theoretical analysis.

**Methods**

NbIrTe4 Sample Preparation

Single crystals of $NbIrTe_4$ were synthesized via the self-flux method. Nb powder (Alfa, 99.99%), Ir powder (Alfa, 99.95%), and Te shot (99.999%) were sealed in a fused silica ampoule under $10^{-6}$ Torr of vacuum. The reagents were mixed to yield a solution of 5 at.% $NbIrTe_4$ in Te. The samples were heated to 1000 C at a rate of 200 C/h, soaked at 1000 C for 24 h, and subsequently cooled to 500 C at 2C/h. Molten Te was centrifuged at 500 C to isolate crystals. Crystals present as thin, silver flakes with a metallic luster. Dimensions of 1 mm × 2 mm × 0.1 mm can be achieved under these conditions.

Device Preparation

A small piece from the original single crystal was exfoliated using Nitto adhesive tape. The thin flakes (typically 100 nm thick) were then transferred on a Si/SiO$_2$ substrate. The electrical contacts were patterned on two ends of the selected flake using photolithography. To obtain good ohmic contacts, we etched the sample in a HF(1):H$_2$O(7) solution for 10 seconds followed by 30 seconds rinse with DI water. Metalization of 20 nm Ti followed by 300 nm Al was done using ebeam evaporation. After lift-off, the device was confirmed to have ohmic contacts using a probe station before mounting to a gold chip carrier and wire bonding gold electrodes to the device. The chip carrier was kept in vacuum for all measurements.

Circularly Polarized Reflectivity and Transient Reflectivity Measurements

A small piece from the original single crystal was exfoliated using Nitto adhesive tape. The thin flakes (typically 100 nm thick) were then transferred on a Si/SiO$_2$ substrate for optical measurements. The Mid-IR laser output from a Ti-Sapphire pumped Optical Parametric Oscillator can be tuned from 0.3 eV to 1.0 eV and is focused onto a single nanoflake using a 40X 0.5 N.A. reflective objective. The reflected light is detected using a LN2 cooled InSb detector. The polarization of the incident light is varied between left- and right-circularly polarized light at 50 kHz using a Hinds Photoelastic Modulator. The change in the reflectivity between left and right circularly polarized light (the Circular Dichroism, CD) can be measured by using a lockin amplifier which is set to 50 kHz.

Transient reflectivity is measured by photoexciting a single nanoflake with a 1.5 eV pump pulse generated from the Ti-Sapphire pump laser, and then measuring the *change* in the CD signal as a function of delay of the probe pulse. The change in the CD signal is measured by taking the output of the first lockin referenced to 50 kHz and using a second lockin which is referenced to an optical chopper which turns on and off the 1.5 eV pump beam.

Photothermoelectric Measurements

The mid-IR probe beam is focused onto the wire-bonded device using the reflective microscope objective. An SRS Current Source is set to 0 (open circuit conditions) and the photovoltage is measured for the probe on and off using an optical chopper and lockin amplifier. Spectra are taken as a function of wavelength. A photovoltage map is taken by raster scanning the focused laser using an Attocube xy piezoelectric scanner.

CPGE Measurements

The mid-IR probe beam is focused onto the wire-bonded device using the reflective microscope objective. With the SRS current source set to 0 (open circuit conditions) the photovoltage is measured for the pump on and off using an optical chopper and lockin amplifier. A variable liquid crystal retardation plate (VLCR) is set to a quarter wave and the waveplate is rotated through 360 degrees so that the polarization can be modulated between left- and right-circularly polarized light and for light which is polarized perpendicular to the contact. The VLCR is calibrated for every wavelength, and left- and right-circular polarizations confirmed by using a wire-grid analyzer.

Supplementary:

S1. The modulation of the photoresponse with linearly polarized light
At 0.69 eV (1800 nm)

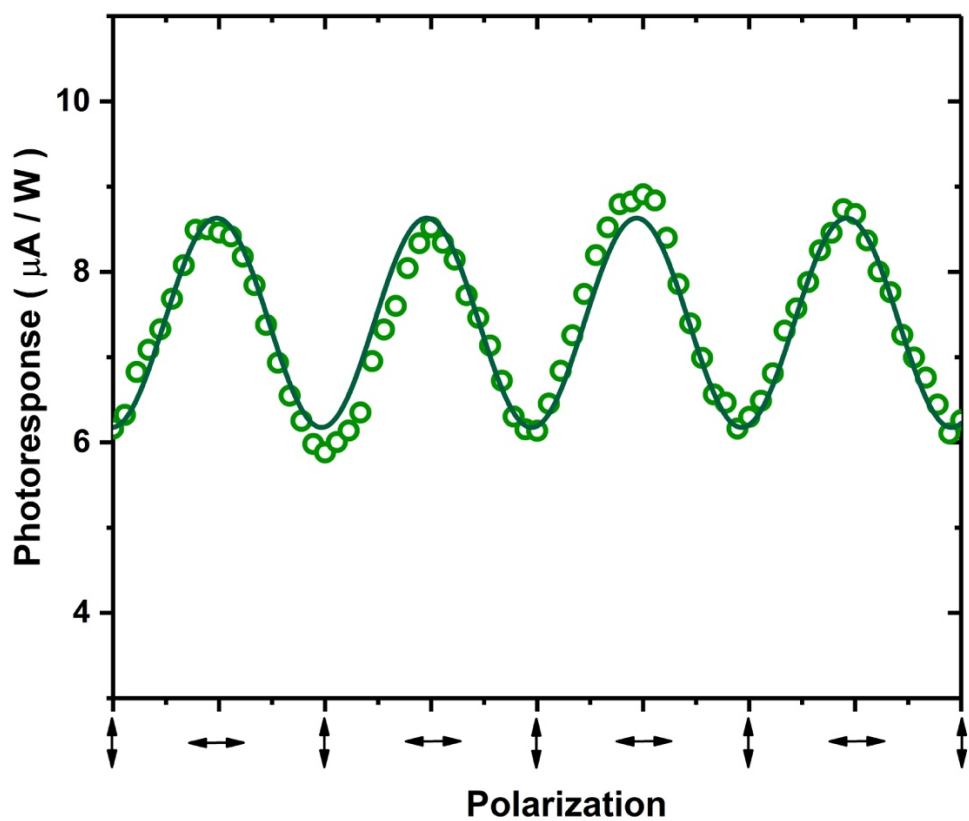

*S 1. Photoresponse as a function of linear polarized light orientation at 0.69 eV.*

**S2. Optical conductivity vs. energy of excitation** (It is good to have them in a single graph)

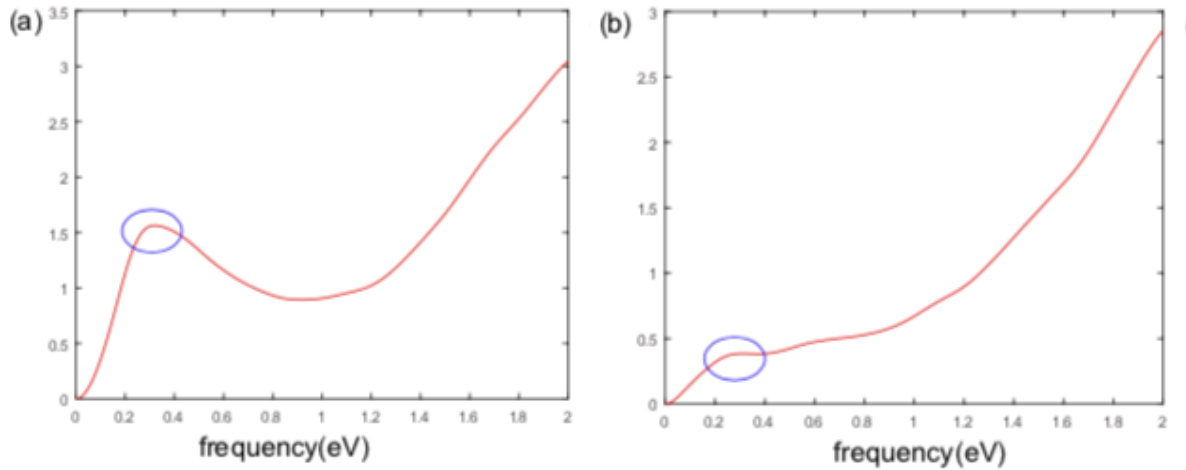

*S 2. Optical conductivity along (a) a-axis and (b) along b-axis of NbIrTe4 crystal versus energy.*

**S3. Power dependent photoresponse to the polarized light.**

The fitting parameters C (circular photogalvanic), L (linear photogalvanic) and D (photothermoelectric) were presented in the main text for various energy excitations. Here we show the power dependence of those parameters for two energy values : 0.41 eV (3000 nm) and 0.34 eV (3600 nm). In both energies the parameters change linearly with the laser power.

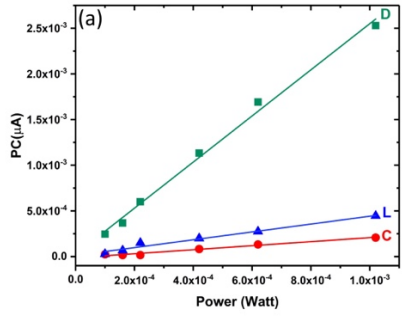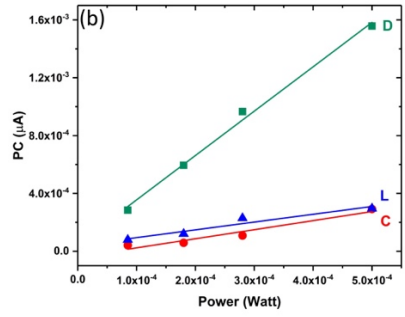

*S 3. Fitting parameters ( C, L and D) vs. laser power for (a) 0.41 eV (3000 nm) and (b) 0.34 eV (3600 nm).*